\documentclass[11pt]{amsart}
\usepackage{hyperref}
\begin{document}

\title[]{A Rule for the Equilibrium of Forces in the Hermitian
Theory of Relativity}
\author{S. Antoci}
\address{Dipartimento di Fisica ``A. Volta'', Via A. Bassi 6,
I-27100 Pavia, Italy}

\thanks{Annalen der Physik, {\bf 44} (1987) 127.}

\begin{abstract}
When the behaviour of the singularities, which are used to
represent masses, charges or currents in exact solutions to the
field equations of the Hermitian theory of relativity, is
restricted by a no-jump rule, conditions are obtained, which
determine the relative positions of masses, charges and currents.
Due to these conditions the Hermitian theory of relativity appears
to provide a unified description of gravitational, colour and
electromagnetic forces.\par\bigskip\noindent
{\bf Eine Regel f\"ur das Gleichgewicht der Kr\"afte in der
Hermi\-teschen Relativit\"atstheorie}\par\smallskip\noindent
Inhalts\"ubersicht. Wenn das Verhalten der Singularit\"aten, die als
Repr\"a\-sentanten von Massen, Ladungen und Str\"omen in exacten
L\"osungen der Feldgleichungen der Hermiteschen
Relativit\"atstheorie benutzt werden, durch eine
Stetigkeitsforderung beschr\"ankt wird, erh\"alt man Bedin\-gungen
f\"ur die relativen Positionen von Massen, Ladungen und Str\"omen.
Entsprechend diesen Bedingungen liefert die Hermitesche
Relativit\"ats\-theorie eine vereinheitlichte Beschreibung der
gravitativen, elektroma\-gnetischen Kr\"afte und Farbkr\"afte.
\end{abstract}
\maketitle

\section{Introduction}
The Hermitian extension of the theory of general relativity, based
on a Hermitian fundamental tensor $g_{ik}=g_{(ik)}+g_{[ik]}$, was
conceived by Einstein in order to provide a unified description of
gravitation and electrodynamics \cite{Einstein}. The study of
solutions of the field equations has shown that the theory
accounts for the existence of gravitational forces, as well as of
forces which do not depend on the distance between charges which
cannot exist as individuals \cite{ Treder1957, Antoci1984}.\par
In the present paper, a rule for the regularity of the exact
solutions is proposed, which confirms the previous findings and
shows that, when the imaginary part $g_{[ik]}$ of the fundamental
tensor fulfills Maxwell's equations, the forces of electrodynamics
between charges and between currents are present too.
\section{A restrictive rule for singularities}

The field equations of the Hermitian theory of relativity are
\cite{Einstein}
\begin{eqnarray}\label{1}
g_{ik,l}-g_{nk}\Gamma^n_{il}-g_{in}\Gamma^n_{lk}=0,\\\label{2}
(\sqrt{-g}g^{[is]})_{,s}=0,\\\label{3}
R_{(ik)}(\Gamma)=0,\\\label{4}
R_{[ij],k}(\Gamma)+R_{[ki],j}(\Gamma)+R_{[jk],i}(\Gamma)=0,
\end{eqnarray}
where $g_{ik}$ is the previously mentioned fundamental tensor, and
$\Gamma^i_{kl}$ is an affine connection, which is Hermitian with
respect to the lower indices, while $R_{ik}(\Gamma)$ is the Ricci
tensor
\begin{equation}\label{5}
R_{ik}=\Gamma^a_{ik,a}-\Gamma^a_{ia,k}
-\Gamma^a_{ib}\Gamma^b_{ak}+\Gamma^a_{ik}\Gamma^b_{ab};
\end{equation}
a comma indicates ordinary differentiation.\par
    Solutions of the field equations depending on two and three
co-ordinates are known; when singularities are allowed in Eqs.
(\ref{3}) and (\ref{4}), some of them appear to represent the
static field of particles endowed with masses and with colour
charges \cite{Antoci1984}. These solutions do not result to be
singular only at the positions where the particles are located;
they display also a lack of elementary flatness on lines
stretching between the particles; the latter singularities
disappear for such configurations of masses and charges, that one
is led to infer that the Hermitian theory of relativity properly
accounts for the existence of gravitational and colour forces.\par
    When singularities in Eqs. (\ref{2}) are allowed, the field
equations of the theory admit solutions in which $g_{[ik]}$
fulfills Maxwell's equations, that clearly represent the
electromagnetic fields associated with point charges, or with
currents running on wires. In this case, however, no extra
singularities in the expected form of deviations from elementary
flatness are encountered, and the conclusion was drawn
\cite{Antoci1984} that the electromagnetic field is there, but it is
dynamically inactive. This conclusion is however unwarranted,
since a more stringent regularity criterion exists, and its
application to the known solutions, while confirming the previous
results on the gravitational and colour forces, shows that the
Hermitian theory of relativity accounts also for the
electromagnetic forces.\par
    The rule is the following: imagine that a solution is given,
in which $g_{ik}$ displays $n$ point - or line - singularities,
that we intend to interpret as masses, charges or currents,
located at different positions in ordinary space. We consider such
an interpretation as allowed only when each of these singularities
does not contain a contribution, in the form of a jump, either
finite of infinite, arising from the presence of the other masses,
charges or currents. Obviously, such a rule has no pretense of
rigour and generality; we consider it just as a heuristic tool,
with which we want to investigate the known solutions.
\section{Consequences of the Restrictive Rule}
    First of all, let us see what happens when the restrictive
criterion proposed above is applied to a Curzon solution
\cite{Curzon} for $n$ masses. Written in canonical cylindrical
co-ordinates $x^1=r$, $x^2=z$, $x^3=\varphi$, $x^4=ct$, $g_{ik}$
takes the Weyl-Levi Civita form \cite{WLC}
\begin{equation}\label{6}
g_{ik}=\left(\begin{array}{cccc}
  -\exp[2(\nu-\lambda)] & 0 & 0 & 0 \\
  0 & -\exp[2(\nu-\lambda)] & 0 & 0 \\
  0 & 0 & -r^2\exp[-2\lambda] & 0 \\
  0 & 0 & 0 & \exp[2\lambda]
\end{array}\right),
\end{equation}
where
\begin{eqnarray}\label{7}
\lambda=-\sum^n_{q=1}\frac{m_q}{p_q},
\\\nonumber
\nu=-\sum^n_{q=1}\frac{m_q^2 r^2}{2p_q^4}
+\sum^{n (q\neq q')}_{q,q'=1}
\frac{m_qm_{q'}}{(z_q-z_{q'})^2}
\left(\frac{r^2+(z-z_q)(z-z_{q'})}{p_qp_{q'}}-1\right);\nonumber
\end{eqnarray}
$m_q$ and $z_q$ are constants, while
\begin{equation}\label{8}
p_q=[r^2+(z-x_q)^2]^{1/2}.
\end{equation}
Such a solution presents singularities, which we intend to
interpret as masses, at $r=0$, $z=z_q$, but it does not satisfy
the restrictive rule proposed above, due to the presence in $\nu$
of terms like
\begin{equation}\label{9}
\mathcal{D}_q=m_q\sum^n_{q'\neq q}
\frac{m_{q'}}{(z_q-z_{q'})^2}
~\frac{r^2+(z-z_q)(z-z_{q'})}{p_qp_{q'}}.
\end{equation}

Imagine that we cross the $q$-th singularity along a given line;
when the co-ordinates of the running pont differ very few from the
co-ordinates of the $q$-th singularity we have
\begin{equation}\label{10}
\mathcal{D}_q\cong m_q\frac{z-z_q}{p_q}
\sum^n_{q'\neq q}m_{q'}\frac{z_q-z_{q'}}{\vert{z_q-z_{q'}}\vert^3}.
\end{equation}
Therefore the term $\mathcal{D}_q$, and hence $g_{ik}$, presents a
finite jump at the position of the $q$-th singularity, unless
\begin{equation}\label{11}
\sum^n_{q'\neq q}m_{q'}\frac{z_q-z_{q'}}
{\vert{z_q-z_{q'}}\vert^3}=0.
\end{equation}

When Eq. (\ref{11}) holds for any value of $q$, i.e. when the
restrictive criterion is fulfilled, even the requirement of
elementary flatness on the $z$-axis is satisfied, and vice-versa;
the no-jump rule, applied to the Curzon solution, provides the
relativistic version of the condition for the equilibrium of $n$
collinear masses at rest.\par
    Let us consider now that solution of the field equations of
the Hermitian theory for which the fundamental tensor, again
referred to cylindrical co-ordinates $r$, $z$, $\varphi$, $ct$,
reads
\begin{equation}\label{12}
g_{ik}=\left(\begin{array}{rrrr}
  -1 & 0 & \delta & 0 \\
  0 & -1 & \varepsilon & 0 \\
  -\delta & -\varepsilon & \zeta & 0 \\
  0 & 0 & 0 & 1
\end{array}\right),
\end{equation}
with
\begin{eqnarray}\label{13}
\delta=i\sum^n_{q=1}\frac{k_q r(z-z_q)}{p_q},\:
\varepsilon=-i\sum^n_{q=1}\frac{k_q r^2}{p_q},
\\\nonumber
\zeta=-r^2\left(1+\sum^n_{q=1}k_q^2
+\sum^{n (q\neq q')}_{q,q'=1}
k_qk_{q'}
\frac{r^2+(z-z_q)(z-z_{q'})}{p_qp_{q'}}\right);\nonumber
\end{eqnarray}
$k_q$ and $z_q$ are constants, while $p_q$ is again defined by Eq.
(\ref{8}), and $i=\sqrt{-1}$. When the sum rule
\begin{equation}\label{14}
\sum^n_{q=1}k_q=0
\end{equation}
is satisfied, the solution, considered in Cartesian co-ordinates,
tends to Minkowski values at space infinity. It displays
singularities at $r=0$, $z=z_q$, which we intend to interpret
\cite{Treder1957,Antoci1984} as colour charges, provided that the
no-jump condition is satisfied at the singularities. This
occurrence is not verified in general, due to the presence in
$\zeta$ of terms like
\begin{equation}\label{15}
\mathcal{D}_q=k_q\sum^n_{q'\neq q}
k_{q'}\frac{r^2+(z-z_q)(z-z_{q'})}{p_qp_{q'}};
\end{equation}
by repeating here the argument used with the Curzon solution, we
find that the no-jump condition is satisfied if
\begin{equation}\label{16}
\sum^n_{q'\neq q}k_{q'}\frac{z_q-z_{q'}}
{\vert{z_q-z_{q'}}\vert}=0.
\end{equation}
When Eq. (\ref{16}) holds for all the values of $q$, elementary
flatness can be ensured on the whole $z$-axis, and vice-versa;
Eq. (\ref{16}) expresses the equilibrium condition of $n$ colour
charges $k_q$ at rest, mutually interacting with forces which do
not depend on the distance \cite{Treder1957,Antoci1984}.\par
    What happens now, when the restrictive condition on
singularities is applied to an electrostatic solution of the
field equations of the Hermitian theory? The solution, written in
Cartesian co-ordinates $x^1=x$, $x^2=y$, $x^3=z$, $x^4=ct$, reads
\cite{Antoci1984}
\begin{equation}\label{17}
g_{ik}=\left(\begin{array}{rrrr}
 -1 &  0 &  0 & a \\
  0 & -1 &  0 & b \\
  0 &  0 & -1 & c \\
 -a & -b & -c & d
\end{array}\right),
\end{equation}
with
\begin{eqnarray}\label{18}
\:a=i\sum^n_{q=1}\frac{h_q (x-x_q)}{p_q^3},\:
b=i\sum^n_{q=1}\frac{h_q (y-y_q)}{p_q^3},\:
c=i\sum^n_{q=1}\frac{h_q (z-z_q)}{p_q^3},
\\\nonumber
d=1-\sum^n_{q=1}\frac{h_q^2}{p_q^4}\\\nonumber
-\sum^{n (q\neq q')}_{q,q'=1}
h_qh_{q'}\frac{(x-x_q)(x-x_{q'})+(y-y_q)(y-y_{q'})
+(z-z_q)(z-z_{q'})}{p_q^3p_{q'}^3};\nonumber
\end{eqnarray}
now
\begin{equation}\label{19}
p_q=[(x-x_q)^2+(y-y_q)^2+(z-z_q)^2]^{1/2},
\end{equation}
while $h_q$, $x_q$, $y_q$ and $z_q$ are constants. We would like to
interpret the singularities of this solution, occurring at
$x=x_q$, $y=y_q$, $z=z_q$, as point electric charges, but we are
prevented to do always so by the restrictive rule, which is not
fulfilled in general, due to the occurrence in $d$ of terms like
\begin{equation}\label{20}
\mathcal{D}_q=h_q\sum^n_{q'\neq q}
h_{q'}\frac{(x-x_q)(x-x_{q'})+(y-y_q)(y-y_{q'})+(z-z_q)(z-z_{q'})}
{p_q^3p_{q'}^3}.
\end{equation}
If we define
\begin{equation}\label{21}
r_{qq'}=[(x_q-x_{q'})^2+(y_q-y_{q'})^2+(z_q-z_{q'})^2]^{1/2},
\end{equation}
we find, in the neighbourhood of the $q$-th singularity
\begin{equation}\label{22}
\mathcal{D}_q\cong h_q\sum^n_{q'\neq q}
h_{q'}\frac{(x-x_q)(x_q-x_{q'})
+(y-y_q)(y_q-y_{q'})+(z-z_q)(z_q-z_{q'})}
{p_q^3r_{qq'}^3}.
\end{equation}
The no-jump condition then requires
\begin{equation}\label{23}
\sum^n_{q'\neq q}h_{q'}\frac{x_q-x_{q'}}{r_{qq'}^3}
=\sum^n_{q'\neq q}h_{q'}\frac{y_q-y_{q'}}{r_{qq'}^3}
=\sum^n_{q'\neq q}h_{q'}\frac{z_q-z_{q'}}{r_{qq'}^3}
=0
\end{equation}
for any value of $q$. Eqs. (\ref{23}) are just the equilibrium
conditions for $n$ point charges $h_q$ at rest, according to
electrostatics.\par
    We now look at the electromagnetic solution \cite{Antoci1984}
for which the fundamental tensor, expressed in Cartesian
co-ordinates $x$, $y$, $z$, $ct$, reads
\begin{equation}\label{24}
g_{ik}=\left(\begin{array}{rrrr}
 -1 &  0 &  e & 0 \\
  0 & -1 &  f & 0 \\
 -e & -f &  h & 0 \\
  0 &  0 &  0 & 1
\end{array}\right),
\end{equation}
with
\begin{eqnarray}\label{25}
e=i\sum^n_{q=1}\frac{l_q (x-x_q)}{p_q^2},\:
f=i\sum^n_{q=1}\frac{l_q (y-y_q)}{p_q^2},
\\\nonumber
h=-1-\sum^n_{q=1}\frac{l_q^2}{p_q^2}
-\sum^{n (q\neq q')}_{q,q'=1}
l_ql_{q'}\frac{(x-x_q)(x-x_{q'})+(y-y_q)(y-y_{q'})}
{p_q^2p_{q'}^2},\nonumber
\end{eqnarray}
where now
\begin{equation}\label{26}
p_q=[(x-x_q)^2+(y-y_q)^2]^{1/2},
\end{equation}
and $l_q$, $x_q$, $y_q$ are arbitrary constants. We intend to
interpret this solution as describing the magnetic field due to $n$
wires parallel to the $z$ axis, on which steady currents $l_q$ are
running, but we can not do so in general, due to the occurrence in
$h$ of terms like
\begin{equation}\label{27}
\mathcal{D}_q=l_q\sum^n_{q'\neq q}
l_{q'}\frac{(x-x_q)(x-x_{q'})+(y-y_q)(y-y_{q'})}
{p_q^2p_{q'}^2}.
\end{equation}
Let us set
\begin{equation}\label{28}
d_{qq'}=[(x_q-x_{q'})^2+(y_q-y_{q'})^2]^{1/2};
\end{equation}
then, in the neighbourhood of the $q$-th wire we have
\begin{equation}\label{29}
\mathcal{D}_q\cong l_q\sum^n_{q'\neq q}
l_{q'}\frac{(x-x_q)(x_q-x_{q'})+(y-y_q)(y_q-y_{q'})}
{p_q^2d_{qq'}^2}.
\end{equation}
$\mathcal{D}_q$, and hence $g_{ik}$, displays an infinite jump
when the $q$-th wire is crossed, unless
\begin{equation}\label{30}
\sum^n_{q'\neq q}l_{q'}\frac{x_q-x_{q'}}{d_{qq'}^2}
=\sum^n_{q'\neq q}l_{q'}\frac{y_q-y_{q'}}{d_{qq'}^2}
=0
\end{equation}
for any value of $q$. Eqs. (\ref{30}) are just the equilibrium
conditions for $n$ parallel wires at rest, run by steady currents
$l_q$, according to electrodynamics.

\section{Conclusion}
The exact solutions and the restrictive rule for singularities
considered in this paper suggest that Einstein's Hermitian theory
of relativity provides a unified description of gravodynamics,
chromodynamics and electrodynamics.\bigskip

Aknowledgements. I am particularly indebted to Professor
H.-J. Treder for continuous support and advice.

\bibliographystyle{amsplain}

\end{document}